\documentclass[prc, showpacs, twocolumn, floatfix]{revtex4}
\usepackage{graphicx}
\usepackage{amsmath, amsfonts, amssymb, bm}
\begin{document}
\title{Laser-assisted nuclear photoeffect}
\author{Anis Dadi$^1$}
\author{Carsten M\"uller$^{1,2}$}
\affiliation{$^1$Max-Planck-Institut f\"{u}r Kernphysik, Saupfercheckweg 1, D-69117 Heidelberg, Germany\\
$^2$Institut f\"ur Theoretische Physik I, Heinrich-Heine-Universit\"at D\"usseldorf, Universit\"atsstr.~1, 40225 D\"usseldorf, Germany}
\date{\today}
\begin{abstract}
Proton emission from nuclei via the nuclear photoeffect in the combined electromagnetic fields of a $\gamma$-ray photon and an intense laser wave is studied. An $S$-matrix approach to the process is developed by utilizing methods known from the theory of nonperturbative laser-atom interactions. As a specific example, photo-proton ejection from halo nuclei is considered. We show that, due to the presence of the laser field, rich sideband structures arise in the photo-proton energy spectra. Their dependence on the parameters and relative orientation of the photon fields is discussed.
\end{abstract}
\maketitle

\section{Introduction}

Studies of the nuclear photoeffect have provided important insights into the structure of the nucleus. The research field was opened with the experiments by Chadwick and Goldhaber on the photodisintegration of the deuteron where $\gamma$-rays from a radioactive source were utilized \cite{Chadwick}. Systematic investigations of the nuclear photoeffect were carried out by Bothe and Gentner in Heidelberg which relied on high-energy $\gamma$-rays produced with the aid of an accelerator \cite{Bothe}. Modern experiments on photonuclear reactions typically apply energetic photons from synchrotron radiation, electron bremsstrahlung or Compton backscattering sources \cite{Nair}.

Due to an enormous and still ongoing technological progress during the last two decades, intense photon beams from powerful laser sources are emerging nowadays into a novel tool for photonuclear studies \cite{Ledingham2003}. For example, a facility devoted particularly to laser-nuclear physics is currently being constructed as part of the Extreme Light Infrastructure project \cite{ELInuc,ELI}.

Direct interactions of laser fields and nuclei are usually very weak because of the large mismatch between the relevant energy scales. Typically, the laser photon energy as well as the electric work performed by the laser field over the nuclear extension are by orders of magnitude smaller than the  nuclear level spacing \cite{Matinyan}. Indirect laser-nucleus coupling schemes have therefore mostly been considered which are mediated by secondary particles such as electrons. Prominent examples are nuclear reactions in laser-produced plasmas \cite{Ledingham2003,Pretzler} where, in particular, photofission and photoneutron production through high-energy bremsstrahlung by laser-accelerated electrons have been observed \cite{photonuclear}. 
Besides, theoreticians have investigated laser-assisted internal conversion \cite{IC}, nuclear reactions via electron-bridge mechanisms \cite{Matinyan,NEET}, and nuclear Coulomb excitation in laser-driven atoms \cite{Atif}. A possibility of lifetime measurements on short-lived excited levels in proton-rich isotopes via laser-driven streaking of emitted protons has been proposed \cite{streaking}. In view of the ever increasing available laser intensities as well as frequencies also prospects for direct laser-induced nuclear reactions are nowadays being explored. Resonant photoexcitation of nuclear transitions may occur when an intense x-ray laser pulse interacts with a counterpropagating nuclear beam of moderately relativistic energy \cite{Thomas}. Laser-induced dynamic nuclear Stark shifts \cite{Thomas2} and excitation of nuclear giant dipole resonances \cite{weide} have been examined as well.

Apart from laser-induced reactions, also laser-assisted processes are of interest. These are processes which can already occur without the presence of a laser field but may be modified when a laser field is present. Examples are the laser-assisted internal conversion \cite{IC} and the laser-assisted nuclear excitation by electron transition \cite{NEET} mentioned above. In general, the requirements on the laser field parameters in order to cause a sizeable effect are substantially less demanding for laser-assisted than for direct laser-induced nuclear processes. 

Laser-assisted processes have been studied thoroughly in atomic phyics. They comprise, for instance, laser-assisted scattering of x-rays, electrons, and ions on atomic targets \cite{Kaminski_Review,ion-atom,ABV_Compton}. In particular, the laser-assisted photoelectric effect in atoms has been investigated intensively \cite{Kalman}. Characteristic modifications of the photoelectron spectra due to the presence of the laser field have been found. Corresponding experimental studies \cite{Glover} have been rendered feasible in recent years by the availability of synchronized optical and extreme-ultraviolet (XUV) beams, with the latter being produced by free-electron lasers or high-harmonic generation. 
The inverse of the laser-assisted atomic photoeffect is laser-assisted radiative recombination of electrons with ions \cite{LARR}.

In the present paper we consider the laser-assisted nuclear photoeffect. That is, photo-proton emission from a nucleus which is subject to the combined electromagnetic fields of a $\gamma$-photon and an intense laser beam. We assume that the $\gamma$-photon energy exceeds the proton separation energy, $\omega_\gamma>E_b$, and study modifications of the $\gamma$-induced proton emission in the presence of an intense laser field of relatively low frequency, $\omega_0\ll E_b$. Regarding the nuclear species, we focus our consideration on one-proton halo nuclei because they possess low proton separation energies \cite{boron,halostate2,halostate,other_halos}. Moreover, their structure allows us to develop a theoretical treatment of the photo-proton emission which is similar to laser-assisted photoionization in atoms \cite{Kalman,LARR}. Symbolically, the process under investigation may be written as 
\begin{equation}
\label{eq:halo_photo_reaction}
^A_Z[{\rm Xp}]+\omega_\gamma + n\omega_0~~\longrightarrow~~^{A-1}_{Z-1}[\mathrm{X}]+^1_1\mathrm{p},
\end{equation} 
where $n$ denotes the number of laser photons involved. We shall show that the assistance by the laser field may have a substantial impact on the photo-proton energy spectra, which are significantly broadened and obtain a rich structure.

The paper is organized as follows. In Sec.~II we derive the quantum mechanical amplitude for the
laser-assisted photoeffect in halo nuclei. The strong-field approximation
will be used, describing the emitted photo-proton in the laser field by a
Volkov state. An analytical expression for the cross section of the
process is given which involves a summation over the number $n$ of
participating laser photons. Our numerical results are presented in
Sec.~III which demonstrate the characteristic influence exerted by the
assisting laser field on the photo-proton emission. The physical origin of
the various effects is discussed and their dependences on the parameters 
of the photon fields and their relative orientation is analysed. 
We conclude with a brief summary and outlook in Sec.~IV.

Natural units with $\hbar = c = \varepsilon_0 = 1$ are used throughout unless otherwise stated.

\section{Theoretical framework}

In this section, we develop a theoretical model to describe the nuclear photoeffect by an incident $\gamma$-photon in the presence of a background laser field. Our approach is inspired by existing theoretical treatments of laser-assisted photoionization of atoms \cite{Kalman,LARR}. We will restrict the consideration to halo nuclei with a weakly bound outer proton because this exotic nuclear species possesses low proton separation energies. 

The laser field is assumed to be a circularly polarized, monochromatic wave of frequency $\omega_0$. It is described by the time-dependent vector potential
\begin{equation}
\label{A}
\vec{A}_L(t)=A_0 \left[ \cos{(\omega_0 t)}\, \vec{e}_1 + \sin{(\omega_0 t)}\, \vec{e}_2 \right],
\end{equation}
with amplitude $A_0$. The corresponding amplitude of the laser electric field is $F_0=\omega_0 A_0$. The unit vectors $\vec{e}_j$ ($j=1,2$) are orthogonal to each other, $\vec{e}_1\cdot \vec{e}_2=0$. Note that in Eq.~\eqref{A} the dipole approximation has been applied which disregards the spatial field dependence. This approximation is well justified because, for the laser parameters under consideration, the scale of the spatial field variations -- which is set by the laser wavelength $\lambda_0=2\pi/\omega_0$ -- is much larger than both the nuclear size $a_0$ and the laser-driven excursion amplitude $\ell_0\sim eF_0/(m\omega_0^2)$ of the emitted proton in the continuum; $m$ denotes the proton mass. Hence, the proton experiences a field which is quasi-constant in space.

\subsection{Derivation of {\it S}-matrix}

The Hamiltonian describing the evolution of the halo proton in the combined fields of the laser beam, the $\gamma$-photon, and the nuclear core reads 
\begin{eqnarray}
\label{H}
H&=\frac{1}{2m}\left(\hat{\vec p}-e\vec{A}_L(t)-e\vec{A}_\gamma(\vec{r},t)\right)^2+V_{\rm nuc}(r)\,.
\end{eqnarray} 
Here, $\hat{\vec p}=-i\vec\nabla$ is the momentum operator of the proton, $e$ the proton charge, and $V_{\rm nuc}$ the potential of the nuclear core. Besides, the vector potential of the incident $\gamma$-photon with energy $\omega_\gamma$, momentum $\vec{k}_\gamma$, and polarization vector $\vec{\varepsilon}$ is 
\begin{eqnarray}
\label{Agamma}
\vec{A}_\gamma(\vec{r},t)=\sqrt{\dfrac{2\pi}{V_\gamma \omega_\gamma}}\, e^{i\left(\vec{k}_\gamma\cdot\vec{r}-\omega_\gamma t\right)}\,\vec{\varepsilon}\,,
\end{eqnarray}
within the normalization volume $V_\gamma$.
The relevant interaction Hamiltonian contained in Eq.~\eqref{H}, which is responsible for $\gamma$-photon absorption by the nucleus, is given by
\begin{eqnarray}
H_{\rm int} = -\frac{e}{m}\left(\hat{\vec p}-e\vec{A}_L(t)\right)\cdot\vec{A}_\gamma(\vec{r},t)\ .
\end{eqnarray}
It can lead to the ejection of the halo proton from the nucleus into the continuum. The corresponding $S$ matrix is
\begin{eqnarray}
\label{S}
S_{fi} = -i\int_{-\infty}^\infty \langle\Psi_f|H_{\rm int}|\Psi_i\rangle dt \,.
\end{eqnarray}
In order to make progress, several approximations may be employed. First, since the laser field will exert only a minor influence on the halo proton bound to the nuclear core, the initial state of the halo proton may be approximated by a stationary field-free state in the nuclear potential $V_{\rm nuc}$, 
\begin{eqnarray}
\Psi_i(\vec{r},t)\approx \phi_0(\vec{r})\, e^{iE_b t}, 
\end{eqnarray}
with the nuclear binding energy $-E_b$. Assuming that the halo proton is in an $s$-state, the space dependent part can be expressed approximately in  Yukawa form as~\cite{halostate2}
\begin{eqnarray}
\label{phi0}
\phi_0(\vec{r})=\dfrac{c_0}{\sqrt{4\pi}}\,\frac{e^{-\beta r}}{\beta r}\ ,
\end{eqnarray}
with $\beta = \nu \sqrt{2mE_b}$, where $\nu$ is a free parameter of order unity which is adjusted to reproduce the measured value of the halo root-mean-square radius.
The normalization constant, guaranteeing that $\langle\phi_0|\phi_0\rangle = 1$, amounts to $c_0=\sqrt{2\beta^3}$.
We point out that the approximation of the wave function in Eq.~\eqref{phi0} is not unique. But, according to Hartree-Fock calculations \cite{HF}, it is appropriate for describing the main halo properties such as the large spatial extension of the halo-proton density.

Second, when the energy of the ejected proton is relatively high (as compared with the nuclear binding energy), the core potential will have a rather small effect on the final proton state in the continuum. The influence of the laser field on this state, however, may be substantial and cannot be ignored. Consequently, the final state may be approximated by a Volkov state, $\Psi_i(\vec{r},t)\approx \psi_{\vec p}(\vec{r},t)$, which exactly accounts for the interaction of the laser field with the ejected proton. It is a solution to the time-dependent Schr\"odinger equation
\begin{eqnarray}
i\frac{\partial}{\partial t}\psi_{\vec p}(\vec{r},t) = \frac{1}{2m}\left(\hat{\vec p}-e\vec{A}_L(t)\right)^2 \psi_{\vec p}(\vec{r},t)
\end{eqnarray}
and reads
\begin{eqnarray}
\label{volkov}
\psi_{\vec p}(\vec{r},t)=\dfrac{e^{i\vec{p}\cdot\vec{r}}}{\sqrt{V}}\exp{\left\{-\dfrac{i}{2m}\int_{t_0}^t\left[\vec{p}-e\vec{A}_L(t')\right]^2dt'\right\}}.
\end{eqnarray}
Here, $\vec p$ denotes the proton momentum outside the field and $V$ is a normalization volume. Note that the lower integration boundary $t_0$ gives rise to an immaterial constant phase factor which may be dropped.

The approximations applied to describe the initial and final states are known in atomic physics as strong-field approximation (SFA). The $S$-matrix of Eq.~\eqref{S} thus becomes $S_{fi}\approx S_{\vec{p}\,0}$, with
\begin{eqnarray}
\label{SFA}
S_{\vec{p}\,0} &=& \frac{ie}{m} \sqrt{\dfrac{2\pi}{V_\gamma \omega_\gamma}} \int dt \int d^3r\ \psi_{\vec p}^*(\vec{r},t) \nonumber \\
& & \times \left({\vec p}-e\vec{A}_L\right)\cdot \vec{\varepsilon} \, e^{i\left(\vec{k}_\gamma\cdot\vec{r}-\omega_\gamma t\right)} \phi_0(\vec{r})\, e^{iE_b t}\,.
\end{eqnarray}
Before we proceed, we point out that we have disregarded a gauge factor of the form $\exp(-ie\vec{A}_L\cdot \vec{r})$ in the initial state in Eq.~\eqref{SFA} (see, e.g., \cite{ABV_Compton}). This is justified in the present case since we will restrict our consideration to field amplitudes $A_0\lesssim 1$\,MeV, whereas the relevant range of the space integral in Eq.~\eqref{SFA} is limited to $r\lesssim a_0\lesssim 10$\,fm due to the presence of the bound nuclear state. Thus, $|\vec{A}_L\cdot \vec{r}|\lesssim 10^{-2}$ and the gauge factor is practically unity.

\subsection{Analytical evaluation of {\it S}-matrix}
The strong-field approximated $S$-matrix in Eq.~\eqref{SFA} can be evaluated by analytical means. In view of the temporal integral we note that,
by inserting Eq.~\eqref{A} into Eq.~\eqref{volkov} and evaluating the integrals in the exponent, the Volkov state of the photo-proton may be written -- after complex conjugation -- as
\begin{equation}
\label{volkov2}
\psi_{\vec p}^*(\vec{r},t)=\frac{e^{-i\vec{p}\cdot\vec{r}}}{\sqrt{V}}\exp{\left[i\left(\frac{p^2}{2m}+U_{\rm p}\right)t \right]}\, f(t),
\end{equation}
with the time-dependent periodic function
\begin{equation}
\label{f}
f(t)=\exp{\Big[-i\left(\alpha_1\sin{\omega_0 t}-\alpha_2\cos{\omega_0 t}\right)\Big]}.
\end{equation}
Here, we have introduced the abbreviation
\begin{equation}
\label{alpha}
\alpha_j = \dfrac{eF_0}{m\omega_0^2}{p_j}\ ,
\end{equation}
with $p_j = \vec{p}\cdot \vec{e}_j$ ($j=1,2$). Besides, the ponderomotive energy of the photo-proton is given by
\begin{align}
\label{Up}
U_{\rm p}&=\dfrac{e^2F_0^2}{2m\omega_0^2}\,.
\end{align}
It represents the cycle-averaged kinetic energy of the proton in the laser field.

We note that $f(t)$ may also be written as 
\begin{equation}
\label{f2}
f(t)=\exp{\left[-i\alpha\sin{\left(\omega_0 t-\eta_0 \right)} \right]},
\end{equation}
with $\alpha=\sqrt{\alpha_1^2+\alpha_2^2}$ and $\eta_0=\arctan(\alpha_2/\alpha_1)$. By exploiting the Jacobi-Anger identity \cite{AS}, we can expand $f(t)$ into Fourier series
\begin{eqnarray}
\label{fourier}
f(t) = \sum_{n=-\infty}^{+\infty}B_n\, e^{-in\omega_0 t}\,,
\end{eqnarray}
with the Fourier coefficients given by $B_n = J_n(\alpha)e^{in\eta_0}$, where $J_n$ is a Bessel function of the first kind of integer order $n$.
Similarly, we find
\begin{eqnarray}
\label{fourier2}
\cos{(\omega_0 t)}f(t) &=& \sum_{n=-\infty}^{+\infty}C_n\, e^{-in\omega_0 t}\,, \nonumber \\
\sin{(\omega_0 t)}f(t) &=& \sum_{n=-\infty}^{+\infty}D_n\, e^{-in\omega_0 t}\,,
\end{eqnarray}
with 
\begin{eqnarray}
\label{CD}
C_n&=\dfrac{1}{2}\left[J_{n+1}(\alpha)e^{i(n+1)\eta_0}+J_{n-1}(\alpha)e^{i(n-1)\eta_0}\right] , \nonumber \\
D_n&=\dfrac{1}{2i}\left[J_{n+1}(\alpha)e^{i(n+1)\eta_0}-J_{n-1}(\alpha)e^{i(n-1)\eta_0} \right].
\end{eqnarray}
Hence, the $S$-matrix \eqref{SFA} adopts the form
\begin{eqnarray}
\label{S3}
S_{\vec{p}\,0} &=& \frac{ie}{m}~\sqrt{\frac{2\pi}{V V_\gamma\omega_\gamma}} 
\sum_{n=-\infty}^{+\infty} \mathcal{M}_n \int d^3r~e^{-i\left(\vec{p}-\vec{k}_\gamma\right)\cdot\vec{r}}\,\phi_0(\vec{r}) \nonumber \\ 
& & \times\, \int dt~e^{i\left(\frac{p^2}{2m}+U_{\rm p}+E_b-\omega_\gamma-n\omega_0\right)t}\,.
\end{eqnarray} 
Here, we introduced
\begin{equation}
\label{Mn}
\mathcal{M}_n = \vec{\varepsilon}\cdot\vec{p}\, B_n - eA_0\varepsilon_1\, C_n - eA_0\varepsilon_2\, D_n,
\end{equation}
with $\varepsilon_j = \vec{\varepsilon} \cdot \vec{e}_j$ ($j=1,2$).

From Eq.~\eqref{S3} it becomes transparent, that the time integral will result in a delta-function which ensures energy conservation in the process. 
Further, the space integral produces the Fourier transform of the bound halo state. By setting ${\vec q}=\vec{p}-\vec{k}_\gamma$, it reads 
\begin{align}
\label{G}
G(q) = \int d^3r~e^{-i\vec{q}\cdot\vec{r}}\,\phi_0(\vec{r}) = \frac{2\sqrt{\pi}\,c_0}{\beta(\beta^2+q^2)}\ .
\end{align}

We finally obtain the following expression for the $S$-matrix
\begin{eqnarray}
\label{S4}
S_{\vec{p}\,0} &=& \dfrac{ie}{m}\,\dfrac{(2\pi)^{3/2}}{\sqrt{V V_\gamma\omega_\gamma}}\sum_{n=n_0}^{+\infty}\mathcal{M}_n\, G(|\vec{p}-\vec{k}_\gamma|) \nonumber\\
& &\times\, \delta\left(\dfrac{p^2}{2m}+U_{\rm p}+E_b-\omega_\gamma-n\omega_0\right),
\end{eqnarray}
where $n_0$ represents the smallest integer that is in accordance with the energy conservation condition. The delta-function fixes, for each laser photon number $n$, the absolute value of the emitted proton momentum at
\begin{equation}
\label{pn}
p_n = \sqrt{2m\left(\omega_\gamma+n\omega_0-U_{\rm p}-E_b\right)}\ .
\end{equation} 
I.e., $n_0$ is the smallest integer which leads to a positive value of the expression under the root.

\subsection{Photonuclear cross section}
The total cross section for the laser-assisted nuclear photo-effect is obtained by squaring the $S$-matrix in Eq.~\eqref{S4}, integrating over the final proton momentum, and dividing out the $\gamma$-photon flux $j=1/V_\gamma$ and a unit time $T$:
\begin{eqnarray}
\label{sigma}
\sigma = \dfrac{1}{jT}\int\dfrac{V~d^3p}{(2\pi)^3}\,|S_{\vec{p}\,0}|^2\ .
\end{eqnarray}
In accordance with the summation over the number of exchanged laser photons in Eq.~\eqref{S4}, also the total cross section decomposes
\begin{eqnarray}
\label{sigma2}
\sigma = \sum_{n=n_0}^{+\infty}\sigma_n
\end{eqnarray}
into a sum over partial cross sections $\sigma_n$. They are given by
\begin{eqnarray}
\label{sigman}
\sigma_n = \dfrac{e^2}{2m\omega_\gamma}\int_0^{\pi} d\theta_p\int_0^{2\pi} d\varphi_p~p_n~\mathcal{M}_n^2~G(|\vec{p}_n-\vec{k}_\gamma|)^2
\end{eqnarray}
with $\vec{p}_n = p_n(\sin\theta\cos\phi,\sin\theta\sin\phi,\cos\theta)$. Due to the energy-conserving delta-function in Eq.~\eqref{S4}, the computation of the partial cross section reduces to an integration over the emission angles of the photo-proton, which can be carried out by numerical means.

\section{Numerical results and Discussion}

Based on Eqs.~\eqref{sigma2} and \eqref{sigman}, we have performed numerical calculations of the laser-assisted nuclear photoeffect in halo nuclei. In what follows, we will consider throughout the one-proton halo isotope $^8$B \cite{boron,halostate2,halostate}. It has a lifetime of about 770~ms and may, thus, be considered stable on the femtosecond to nanosecond time scales of intense laser pulses. The proton separation energy is very low and amounts to $E_b=137$~keV \cite{halostate2}. For the root-mean-square distance in the $^7\mathrm{Be}$-proton system we employ the value $R_{\rm rms}\approx 4.73$~fm~\cite{halostate} and, accordingly, set $\nu = 1.84$ in Eq.~\eqref{phi0}.

The incident $\gamma$-photon is assumed to have a relatively high energy of $\omega_\gamma=3$~MeV in order to guarantee that the nuclear core potential $V_{\rm nuc}$ in Eq.~\eqref{H} has a rather small influence on the proton in the continuum. The $\gamma$-photon propagates into the $x$ direction and is polarized along the $z$ axis, $\vec{ \varepsilon} = \vec{e}_z$. We note that $\gamma$-photons of several MeV energy can be produced today through bremsstrahlung of laser-accelerated electron bunches \cite{GammaSource}. An experimental setup to probe the laser-assisted nuclear photoeffect could therefore rely on two sources of intense laser radiation: one to generate the $\gamma$-rays and the second one to provide the assisting laser field.

Several laser frequencies in the XUV and x-ray domains with various intensities have been applied in order to reveal the dependence of the process on these parameters. The polarization vectors of the laser field have almost always been chosen as $\vec{e}_1 = \vec{e}_y$ and $\vec{e}_2 = \vec{e}_z$ [see Eq.~\eqref{A}]. Hence, the polarization vector of the $\gamma$-photon lies in the polarization plane of the circularly polarized laser field. Additionally, in Section III.C another polarization plane of the laser is considered in order to study the relevance of the field geometry.

We briefly comment on the current status of high-intensity laser sources operating in the XUV and x-ray regimes. The FLASH facility (DESY, Germany), which is based on a free-electron laser, produces brilliant photon beams with XUV frequencies of $\omega_0 \sim 100$\,eV at peak intensities up to $\sim 10^{17}$\,W/cm$^{2}$ \cite{FLASH}. The free-electron laser at the Linac Coherent Light Source (SLAC, Stanford) is presently able to generate x-ray photon beams with $\omega_0 \sim 1$\,keV at $\sim 10^{18}$\,W/cm$^{2}$ \cite{LCLS}. A substantial further increase of the attainable peak intensities is envisaged at both facilities. High-intensity coherent XUV and x-ray pulses can also be created through harmonic emission from laser-irradiated plasma surfaces \cite{surface}.

\subsection{Transition from perturbative to nonperturbative laser-proton coupling}

\begin{figure}[b]
\centering
\rotatebox{-90}{\includegraphics[width=0.3\textwidth]{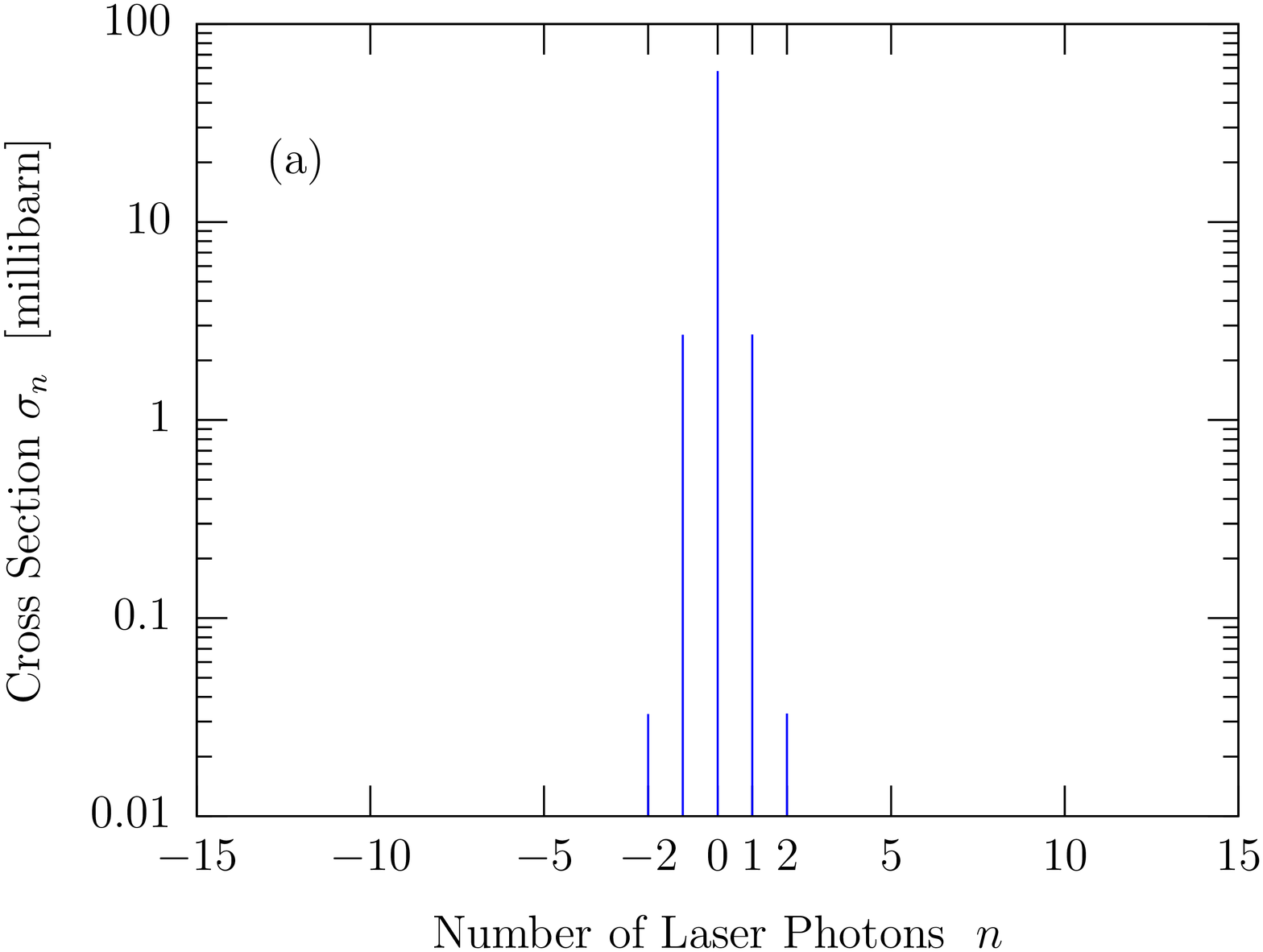}}
\rotatebox{-90}{\includegraphics[width=0.3\textwidth]{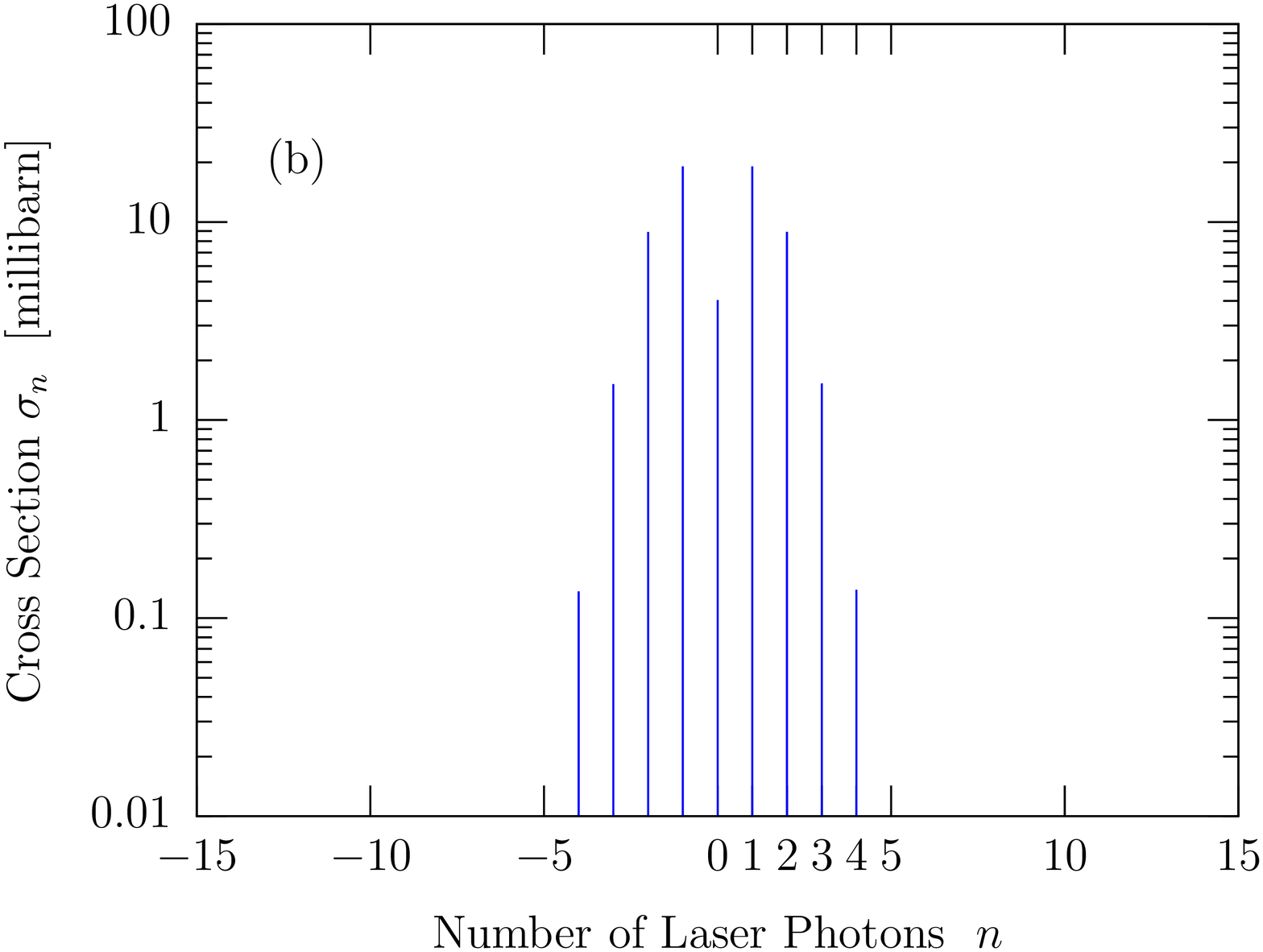}}
\rotatebox{-90}{\includegraphics[width=0.3\textwidth]{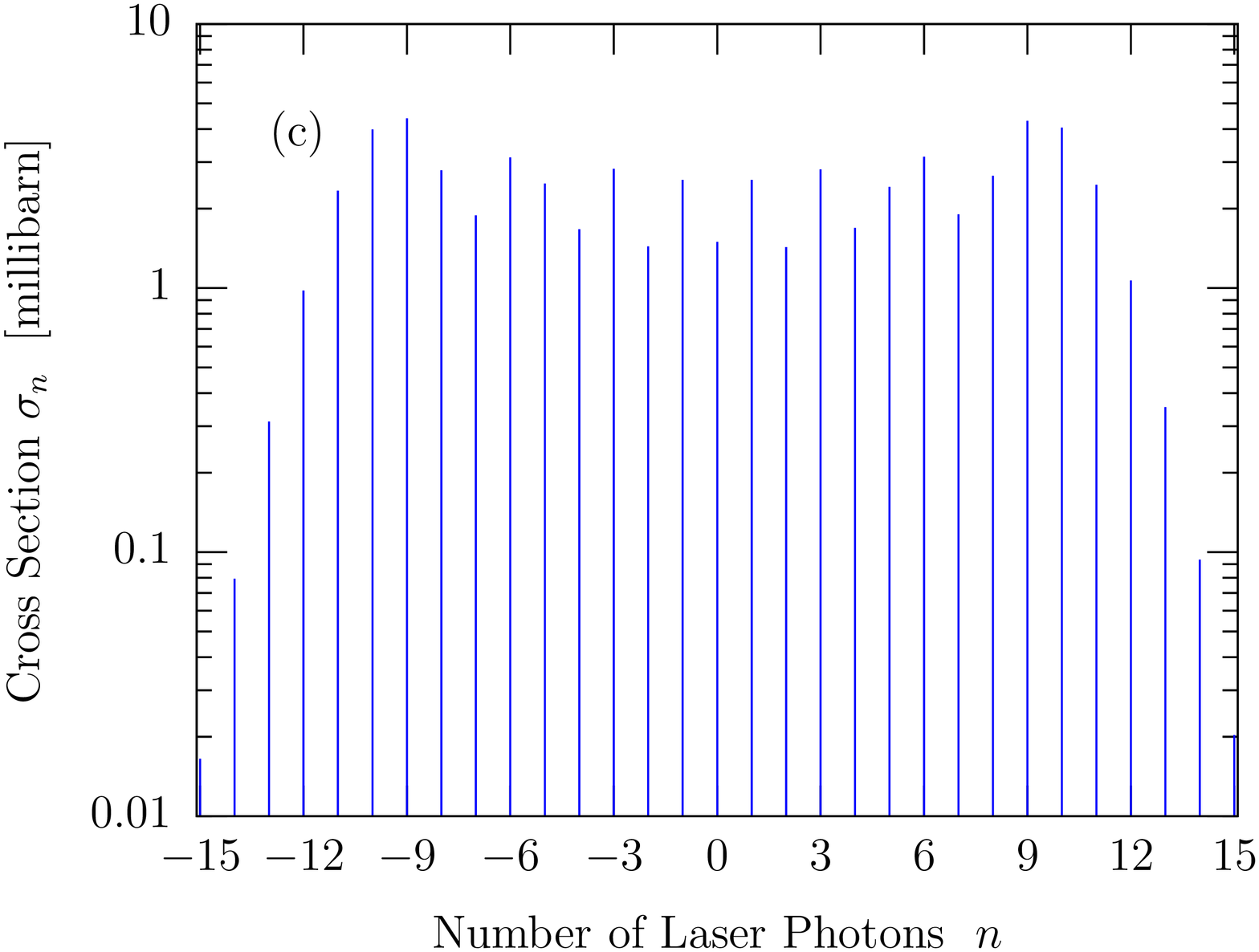}}
\caption{Distributions of the partial cross sections $\sigma_n$ for laser-assisted proton emission from $^8_5$B, as a function of the number of absorbed ($n>0$) or emitted ($n<0$) laser photons. The circularly polarized laser field has a frequency of $\omega_0=2$\,keV. Its intensity amounts to (a) $I = 4.0\times 10^{21}$\,W/cm$^{2}$, (b) $I = 1.0\times 10^{23}$\,W/cm$^{2}$, and (c) $I = 2.5\times 10^{24}$\,W/cm$^{2}$, respectively. The $\gamma$-ray photon has an energy of $\omega_\gamma=3$\,MeV; its polarization vector lies in the polarization plane of the laser.}
\label{pert}
\end{figure}

Figure~\ref{pert} shows the partial cross sections $\sigma_n$ as a function of the number $n$ of emitted ($n<0$) or absorbed ($n>0$) laser photons. The laser frequency lies in the x-ray domain and amounts to $\omega_0=2$~keV. The laser intensity varies between $\sim 10^{21}$--$10^{24}$~W/cm$^2$. Note that the distributions shown in Fig.~\ref{pert} reflect the energy spectra of the emitted proton: at a given value of $n$, the proton energy is $E_{n}=\omega_\gamma + n\omega_0 - U_{\rm p} - E_b$. The fraction of protons with this energy amounts to $\sigma_n/\sigma$.

In the absence of a laser field, the photo-proton energy spectrum would consist of a single line located at $\omega_\gamma - E_b\approx 2.86$~MeV. As Fig.~\ref{pert} illustrates, the presence of the laser field leads to sidebands in the energy distribution which surround the central line and are equally spaced by a laser photon energy $\omega_0$.

When the laser intensity is relatively low, only the first few sidebands show up [see Fig.~\ref{pert}(a)]. Their height, as compared with the central line, is suppressed because the laser-proton interaction represents here a small perturbation only. The expansion parameter of the corresponding perturbation series is given by 
\begin{eqnarray}
\label{alphamax}
\alpha_{\rm max}=\frac{eF_0p_{\rm max}}{m\omega_0^2}
\end{eqnarray}
where $p_{\rm max}$ denotes the maximum value of the photo-proton momentum component in the polarization plane of the laser field. It enters into the partial cross section $\sigma_n$ as maximum argument of the Bessel function $J_n(\alpha)$. Since $J_n(x)\approx \frac{1}{n!}\left(\frac{x}{2}\right)^{n}$ for $x\ll 1$ at $n\ge 0$ and $J_{-n}(x)=(-1)^n J_n(x)$ \cite{AS}, higher photon orders are suppressed when the expansion parameter is small. In particular, the first and second sidebands visible in Fig.~\ref{pert}(a) are reduced by relative factors of $\sigma_{\pm 1}/\sigma_0 \sim \alpha_{\rm max}^2/4\sim 0.05$ and $\sigma_{\pm 2}/\sigma_0 \sim \alpha_{\rm max}^4/8\sim 10^{-3}$, as compared with the central line.

As Fig.~\ref{pert}(b) shows, when the laser intensity is increased, the number of sidebands grows. Besides, the contribution from the sidebands to the total cross section \eqref{sigma2} can be comparable or even exceed the contribution from the central line. When the laser intensity is increased even further, an extended plateau of sidebands of similar height arises [see Fig.~\ref{pert}(c)]. Here, the value of the coupling parameter \eqref{alphamax} has raised to $\alpha_{\rm max} \approx 2.4$, indicating that a transition from perturbative to nonperturbative laser-proton coupling has occured. In terms of the photo-proton energy, the plateau covers the range from about 2.84~MeV to 2.88~MeV. Note that the value of the ponderomotive energy is small, $U_{\rm p}\approx 2$~eV.

We point out that the total cross section in Figs.~\ref{pert}(a)--(c) is always the same, $\sigma = \sum_n \sigma_n \approx 63.4$~mb. Hence, while the presence of the laser field distributes the photo-proton energies over a broad range, it does not affect the total probability of the process. For comparison we note that an estimate of the total cross section based on the Bethe-Peierls formula \cite{LL} yields a value of about 20~mb and, thus, agrees with the prediction from our model by the order of magnitude.

\subsection{Fully nonperturbative laser-proton interaction}

\begin{figure}[t]
\centering
\rotatebox{-90}{\includegraphics[width=0.35\textwidth]{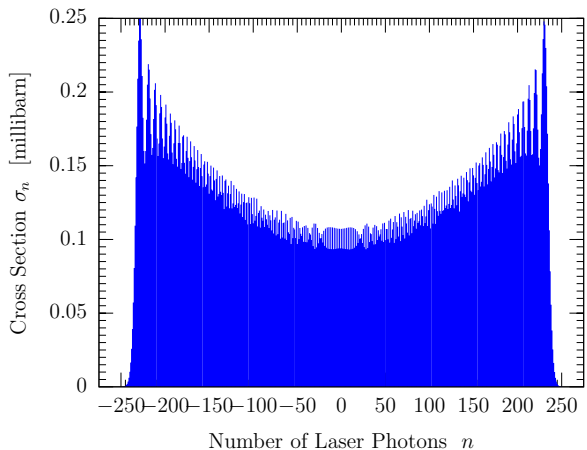}}
\caption{Same as Fig.~\ref{pert}, but for a laser frequency of $\omega_0=100$\,eV at intensity $I = 6.25\times 10^{21}$\,W/cm$^{2}$.}
\label{fully}
\end{figure}

In Section III.A we have seen that the coupling between the emitted proton and the x-ray laser field becomes stronger when the laser intensity is enhanced. By inspection of the coupling parameter \eqref{alphamax} we may expect that the interaction strength grows further when the laser frequency is decreased. This is confirmed in Fig.~\ref{fully} where the distribution of the partial cross sections $\sigma_n$ is displayed at an XUV laser frequency of $\omega_0=100$~eV and a field intensity of $I=6.25\times10^{21}~\mathrm{W/cm^2}$ (and otherwise unchanged parameters). The laser-proton interaction exhibits a highly nonperturbative character now which is related to a large value of the coupling parameter, $\alpha_{\rm max}\approx 238$. A multitude of sidebands appears, forming a quasi-continuous distribution and featuring several maxima whose height is growing towards the side wings. Hence, the $\gamma$-induced proton emission proceeds most likely with the simultaneous absorption or emission of hundreds of 
laser photons. The cutoff of the 
distribution at $|n|\gtrsim 240$ may be traced back to the properties of the Bessel functions $J_n(\alpha)$ of large order which quickly approach zero in the region where $|n|>\alpha$ \cite{AS}. The proton energy varies between 2.84~MeV and 2.88~MeV, accordingly.

The width of the distribution in Fig.~\ref{fully} can be understood by a classical-physics consideration. Let us consider a proton moving freely in an external laser field of the form \eqref{A}. The initial proton momentum outside the field is denoted as $\vec{p}$. Then the classical instantaneous energy of the proton in the field is 
\begin{eqnarray}
E_{\vec{p}\,}(t) = \frac{p^2}{2m} + U_{\rm p} - \frac{eA_0}{m}\left[ p_1 \cos(\omega_0 t) + p_2 \sin(\omega_0 t)\right]
\end{eqnarray}
which shows an oscillatory temporal behavior. The amplitude of these oscillations is largest, when the proton momentum lies in the polarization plane of the laser field. Accordingly, the proton energy can vary within the boundaries
\begin{eqnarray}
\frac{p^2}{2m} + U_{\rm p} - \frac{eF_0p}{m\omega_0} \le E_{\vec{p}\,}(t) \le \frac{p^2}{2m} + U_{\rm p} + \frac{eF_0p}{m\omega_0}\,,
\end{eqnarray}
resulting in an energetic width of $\Delta E =2eF_0p/(m\omega_0)$. The same width of the distribution of photo-proton energies $E_{n}=\omega_\gamma + n\omega_0 - U_{\rm p} - E_b$ results in the quantum description of the process. There, it follows from the maximum number of laser photons involved in the process which is given by $n_{\rm max}\approx \alpha_{\rm max}$. The latter corresponds to a total laser photon energy of $n_{\rm max}\omega_0 \approx eF_0p_{\rm max}/(m\omega_0)$ which is either absorbed from or emitted into the laser field.

Concluding this section, we point out that a photo-proton energy spectrum with a total width of 40\,keV as in Fig.~\ref{fully} would also result from the assistance of an optical laser with a frequency of $\omega_0 = 2$\,eV and an intensity of $I = 2.5\times 10^{18}$\,W/cm$^{2}$. Such field parameters are available today from powerful table-top laser systems. While the general appearance of the proton spectrum will be similar to the one in Fig.~\ref{fully}, its numerical calculation is substantially more involved due to the very large number $n\sim 10^4$ of laser photons participating in the process.

\subsection{Dependence on field geometry} 

\begin{figure}[tbh]
\centering
\rotatebox{-90}{\includegraphics[width=0.35\textwidth]{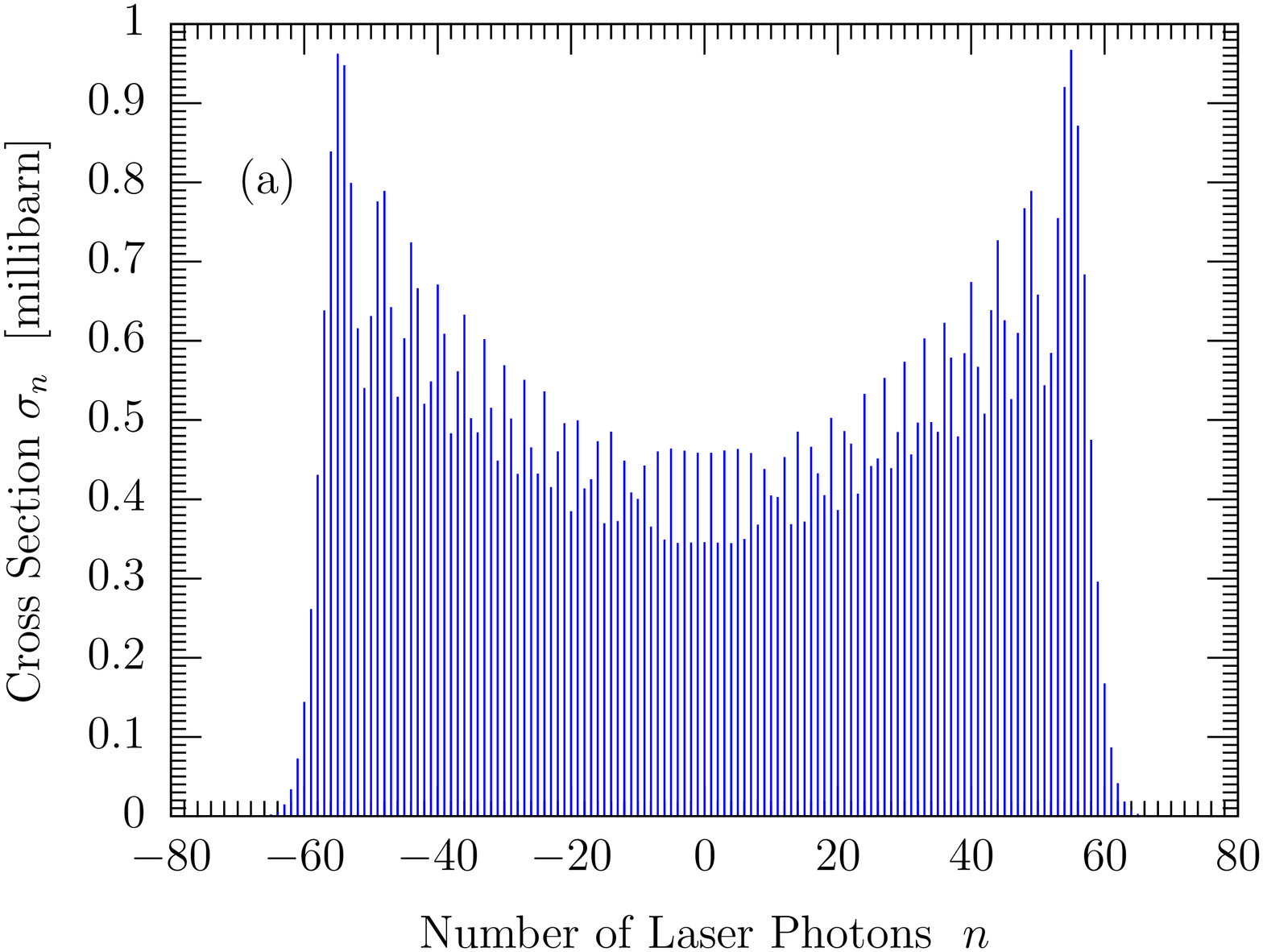}}
\rotatebox{-90}{\includegraphics[width=0.35\textwidth]{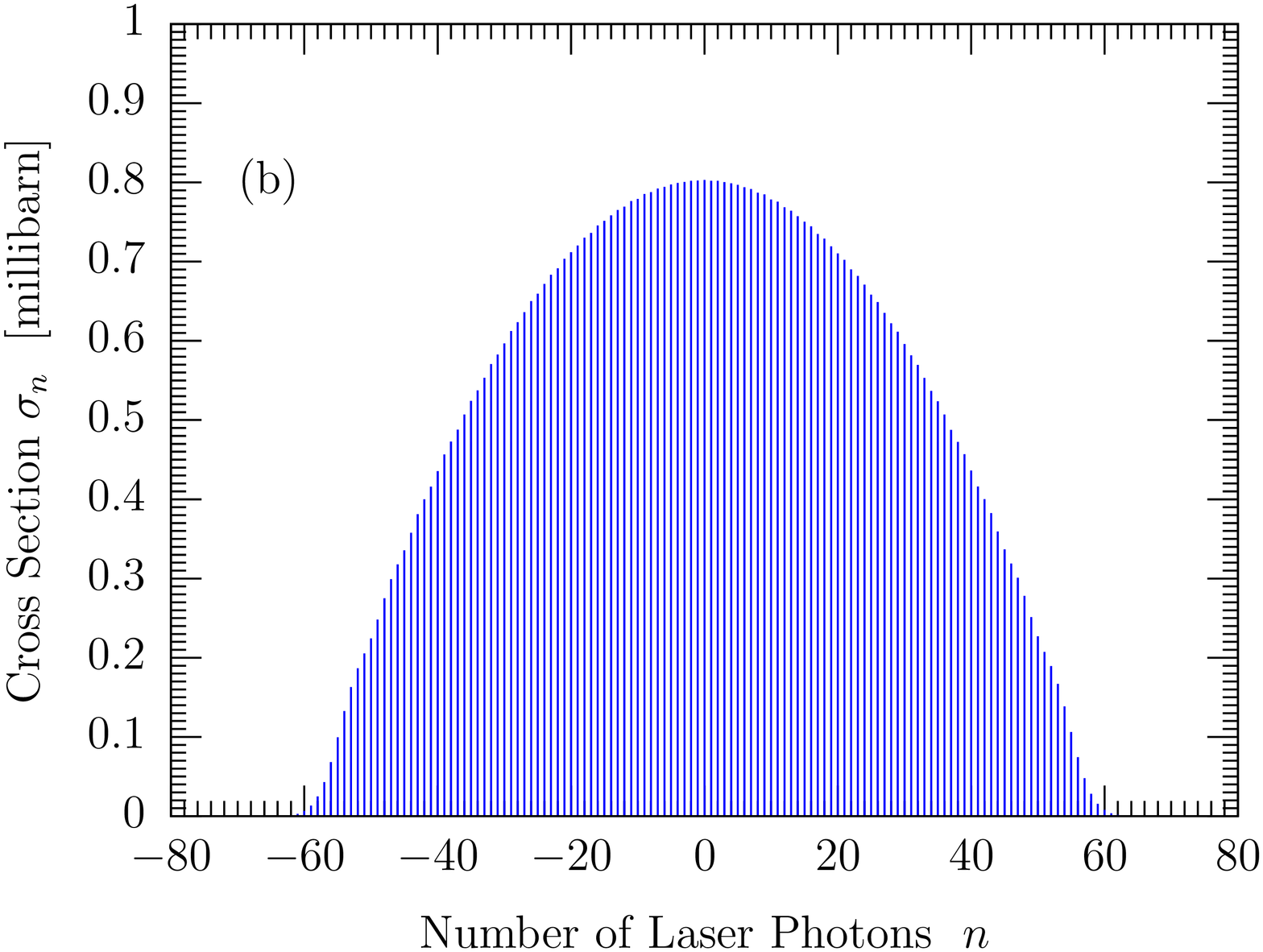}}
\caption{(a) Same as Fig.~\ref{pert}, but for laser frequency $\omega_0=200$\,eV 
and laser intensity $I = 6.25\times 10^{21}$\,W/cm$^{2}$; (b) Same as (a) but the 
laser field is polarized in the $x$-$y$ plane, i.e., perpendicularly to the 
polarization direction of the $\gamma$-photon.}
\label{geo}
\end{figure}

Finally, we address the question how the relative orientation of
the laser field with respect to the polarization direction of the 
$\gamma$-photon influences the proton emission. 
So far we have considered the situation where the $\gamma$-photon polarization 
vector lies in the polarization plane of the circularly polarized laser field 
[in the notation of Eqs.~\eqref{A} and \eqref{Agamma}, 
$\vec{e}_1=\vec{e}_y$, $\vec{e}_2=\vec{\varepsilon}=\vec{e}_z$]. 
Figure~\ref{geo}(a) shows a corresponding distribution of the partial
cross sections $\sigma_n$ for a laser frequency of $\omega_0 = 200$\,eV and 
a laser intensity of $6.25\times 10^{21}~\mathrm{W/cm^2}$. It extends over
an energetic width from about 2.85~MeV to 2.87~MeV.

In contrast, when the $\gamma$-photon polarization is perpendicular to the
polarization plane of the laser ($\vec{e}_1=\vec{e}_x$, 
$\vec{e}_2=\vec{e}_y$, $\vec{\varepsilon}=\vec{e}_z$), with all other
parameters remaining unchanged, the distribution in Fig.~\ref{geo}(b)
results. While the total widths of both distributions are the same, their shapes
are qualitatively different. The distribution in Fig.~\ref{geo}(a)
exhibits a rich structure with maxima at the outer edges, as we have seen
before in Sec.~III.B. Instead, for the field configuration of
Fig.~\ref{geo}(b), the distribution possesses a smooth bell-shaped form
with the maximum lying at the center.

The distinct dependence on the relative field orientation can be
understood by noting the following points:

(i) from the usual photoeffect it is known that, in the nonrelativistic
domain, the photo-proton is preferentially emitted under small angles
around $\theta\approx 0$ with respect to the polarization direction of the
$\gamma$-photon \cite{LL};

(ii) the coupling strength between the proton and the laser field depends
on the proton momentum component within the laser polarization plane [see
Eq.~\eqref{alphamax}].

Therefore, in our notation, the majority of protons is emitted with a large momentum
component $p_z$ (as compared with $p_x$ and $p_y$) and it matters whether
this component lies inside or outside the polarization plane of the laser.
When $p_z$ lies inside the laser polarization plane, then the laser
couples strongly to the proton and the Bessel function argument is, on
average, large ($\alpha\propto \cos\theta$). This renders the emission or
absorption of a large number of laser photons very likely, leading to the
pronounced side wings in Fig.~\ref{geo}(a).

When, instead, the $\gamma$-photon polarization is perpendicular to the
polarization plane of the laser, the laser-proton coupling is mediated
through the relatively small momentum components $p_x$ and $p_y$, so that
the Bessel function argument is mostly small ($\alpha\propto \sin\theta$).
As a consequence, it is more probable that the proton emission proceeds
with participation of only a small number of  laser photons whereas high
photon orders are surpressed [see Fig.~\ref{geo}(b)].

\vspace{0.5cm}

Before we proceed to the conclusion, we note that all calculations of Sec.~III 
have also been performed by applying another form of the bound halo state in order 
to test the sensitivity of our predictions to the particular choice
of the Yukawa wave function in Eq.~\eqref{phi0}. To this end, a ``hydrogen-like'' 
$2s$ state has been employed which is often used to model the neutron halo in 
$^{11}$Be \cite{2s}. Only very small differences were found. For instance, when the
$2s$ state is used, the resulting total cross section deviates from
the value given above by a few percent. Thus, the particular form of the bound 
halo wave function has only a minor influence on our results.

\section{Conclusion and outlook}

Photo-proton emission from halo nuclei subject to the combined fields
of a $\gamma$-photon and an intense, circularly polarized laser wave 
has been investigated. An $S$-matrix theory within the framework of the
strong-field approximation as known from atomic physics was presented and
applied to the laser-assisted photoeffect in $^8$B. It was shown that
the presence of the laser field may substantially influence the energy
distribution of the emitted proton. The transition from perturbative to
nonperturbative laser-proton interaction was illustrated and the fully
nonperturbative regime was discussed. Here, the width of the energy
distribution is determined by the maximum classical kinetic energy which
the proton may gain (or loose) in the laser field. The width therefore
scales with the amplitude of the laser vector potential. For the field
parameters under consideration, the total cross section of the
photo-proton emission is not affected by the presence of the laser field.

In summary, the laser-assisted nuclear photoeffect shares many similarities 
with its atomic-physics counterpart \cite{Kalman,Glover}. We emphasize that, 
while our theory has been designed for application to one-proton halo 
nuclei, qualitatively similar effects can be expected for laser-assisted 
photo-proton emission from ordinary nuclei as well. 

As an outlook, we note that further analogies between atomic and nuclear 
processes in strong laser fields may exist. For example, 
the inverse of the laser-assisted nuclear photoeffect would be
radiative proton capture by a nucleus in the presence of a laser field. 
The corresponding process in atomic physics is laser-assisted radiative 
recombination of electrons with ions \cite{LARR}. This process also
represents the final step in high-harmonic generation from atoms or 
molecules in external laser fields \cite{HHG}. At future high-intensity laser
facilities, corresponding processes could be stimulated in nuclei as well.
At optical laser intensities of $I\gtrsim 10^{24}$\,W/cm$^{2}$
as envisaged at the Extreme Light Infrastructure \cite{ELI}, the
laser-driven proton dynamics would even become relativistic.

\section*{Acknowledgments}
Valuable input by K. Z. Hatsagortsyan is gratefully acknowledged.
We also thank W. N\"ortersh\"auser for useful conversations on halo nuclei.
A.~D. acknowledges the scholarship support from the International 
Max Planck Research School for Quantum Dynamics in Physics, Chemistry and Biology (IMPRS-QD).



\begin{thebibliography}{33}
\bibitem{Chadwick} J. Chadwick and M. Goldhaber, Nature \textbf{134}, 237 (1934).

\bibitem{Bothe}
W. Bothe and W. Gentner, Z. Phys. \textbf{106}, 236 (1937);
\textbf{112}, 45 (1939).

\bibitem{Nair} See, e.g., 
Y. Assafiri \textit{et al.}, Phys. Rev. Lett. \textbf{90}, 222001 (2003);
C. Nair \textit{et al.}, Phys. Rev. C \textbf{78}, 055802 (2008);
R. R\"ohlsberger, K. Schlage, B. Sahoo, S. Couet, and R. R\"uffer, Science \textbf{328}, 1248 (2010).

\bibitem{Ledingham2003} K. W. D. Ledingham, P. McKenna, and R. P. Singhal, Science \textbf{300}, 1107 (2003);
D. Umstadter, J. Phys. D \textbf{36}, R151 (2003);
K. W. D. Ledingham and W. Galster, New J. Phys. \textbf{12}, 045005 (2010).

\bibitem{ELInuc}
N. V. Zamfir, D. Habs, F. Negoita, and D. Ursescu, ``Extreme Light Infrastructure: nuclear physics'', Proc. SPIE 8080, 80800X (2011); doi:10.1117/12.890139

\bibitem{ELI} For current information, see http://www.eli-laser.eu.

\bibitem{Matinyan} S. Matinyan, Phys. Rep. \textbf{298}, 199 (1998).

\bibitem{Pretzler} G. Pretzler \textit{et al.}, Phys. Rev. E {\bf 58}, 1165 (1998);  
T. Ditmire \textit{et al}., Nature (London) \textbf{398}, 489 (1999).

\bibitem{photonuclear}
K. W. D. Ledingham \textit{et al.}, Phys. Rev. Lett. \textbf{84}, 899 (2000);
T. E. Cowan \textit{et al.}, Phys. Rev. Lett. \textbf{84}, 903 (2000);
H. Schwoerer, P. Gibbon, S. D\"usterer, R. Behrens, C. Ziener, C. Reich, and R. Sauerbrey, Phys. Rev. Lett. \textbf{86}, 2317 (2001).

\bibitem{IC} P. K\'alm\'an and J. Bergou, Phys. Rev. C \textbf{34}, 1024 (1986);
P. K\'alm\'an, Phys. Rev. C \textbf{39}, 2452 (1989); 
D. Kis, P. K\'alm\'an, T. Keszthelyi, and J. Sz\'iv\'os, Phys. Rev. A \textbf{81}, 013421 (2010).

\bibitem{NEET}
P. K\'alm\'an and T. Keszthelyi, Phys. Rev. A \textbf{47}, 1320 (1993);
S. Typel and C. Leclercq-Willain, Phys. Rev. A \textbf{53}, 2547 (1996). 

\bibitem{Atif} 
J. C. Solem and L. C. Biedenharn, J. Quant. Spectrosc. Radiat. Transfer \textbf{40}, 707 (1988);
J. F. Berger, D. M. Gogny, and M. S. Weiss, Phys. Rev. A \textbf{43}, 455 (1991);
F. X. Hartmann, D. W. Noid, and Y. Y. Sharon, Phys. Rev. A \textbf{44}, 3210 (1991);
N. Milosevic, P. B. Corkum, and T. Brabec, Phys. Rev. Lett. \textbf{92}, 013002 (2004);
A. S. Kornev and B. A. Zon, Laser Phys. Lett. \textbf{4}, 588 (2007);
A. Shahbaz, C. M\"uller, T. J. B\"urvenich, and C. H. Keitel, Nucl. Phys. A \textbf{821}, 106 (2009).

\bibitem{streaking} D. Habs, T. Tajima, J. Schreiber, C. P. J. Barty, M. Fujiwara, and P. G. Thirolf, Eur. Phys. J. D \textbf{55}, 279 (2009).

\bibitem{Thomas} T. J. B\"{u}rvenich, J. Evers, and C. H. Keitel, Phys. Rev. Lett. \textbf{96}, 142501 (2006);
W.-T. Liao, A. P\'alffy, and C.~H. Keitel, Phys. Lett. B \textbf{705}, 134 (2011).

\bibitem{Thomas2} T. J. B\"{u}rvenich, J. Evers, and C. H. Keitel, Phys. Rev. C {\bf 74}, 044601 (2006).

\bibitem{weide} H. A. Weidenm\"uller, Phys. Rev. Lett. \textbf{106}, 122502 (2011)

\bibitem{Kaminski_Review} F. Ehlotzky, K. Krajewska, and J. Z. Kami{\'n}ski, Rep. Prog. Phys. {\bf 72}, 046401 (2009).

\bibitem{ion-atom} T. Kirchner, Phys. Rev. Lett. \textbf{89}, 093203 (2002); E. L\"otstedt, U. D. Jentschura, and C. H. Keitel, Phys. Rev. Lett. \textbf{101}, 203001 (2008); A. B. Voitkiv, B. Najjari, and J. Ullrich, Phys. Rev. Lett. \textbf{103}, 193201 (2009);
R. Kanya, Y. Morimoto, and K. Yamanouchi, Phys. Rev. Lett. \textbf{105}, 123202 (2010).

\bibitem{ABV_Compton} A. B. Voitkiv, N. Gr\"un, and J. Ullrich, J. Phys. B {\bf 36}, 1907 (2003).

\bibitem{Kalman} See, e.g., P. K{\'a}lm{\'a}n, Phys. Rev. A {\bf 38}, 5458 (1988);
C. Leone, S. Bivona, R. Burlon, and G. Ferrante, Phys. Rev. A \textbf{38}, 5642 (1988);
V. V{\'e}niard, R. Ta\"ieb, and A. Maquet, Phys. Rev. Lett. {\bf 74}, 4161 (1995);
D. B. Milo\v{s}evi{\'c} and F. Ehlotzky, Phys. Rev. A {\bf 57}, 2859 (1998);
C. Buth and R. Santra, Phys. Rev. A {\bf 75}, 033412 (2007);
A. K. Kazansky, A. V. Grigorieva, and N. M. Kabachnik, Phys. Rev. Lett. \textbf{107}, 253002 (2011).

\bibitem{Glover} T. E. Glover, R. W. Schoenlein, A. H. Chin, and C. V. Shank, Phys. Rev. Lett. \textbf{76}, 2468 (1996);
P. Johnson \textit{et al.}, Phys. Rev. Lett. \textbf{95}, 013001 (2005);
M. Meyer \textit{et al.}, Phys. Rev. Lett. \textbf{101}, 193002 (2008);
K. Kl\"under \textit{et al.}, Phys. Rev. Lett. \textbf{106}, 143002 (2011).

\bibitem{LARR} See, e.g., J. Z. Kami{\'n}ski and F. Ehlotzky, J. Mod. Opt. {\bf 50}, 621 (2003);
C. M\"uller, A. B.Voitkiv, and B. Najjari, J. Phys. B {\bf 42}, 221001 (2009).

\bibitem{boron} T. Minamisono \textit{et al.}, Phys. Rev. Lett. \textbf{69}, 2058 (1992); 
L. Trache, F. Carstoiu, C. A. Gagliardi, and R. E. Tribble, Phys. Rev. Lett. \textbf{87}, 271102 (2001); 
D. Cortina-Gil \textit{et al.}, Phys. Lett. B \textbf{529}, 36 (2002); Nucl. Phys. A \textbf{720}, 3 (2003);
T. Sumikama \textit{et al.}, Phys. Rev. C \textbf{74}, 024327 (2006).

\bibitem{halostate2} M. Fukuda \textit{et al.}, Nucl. Phys. A \textbf{656}, 209 (1999).

\bibitem{halostate} M. H. Smedberg \textit{et al.}, Phys Lett. B \textbf{452}, 1 (1999).

\bibitem{other_halos} For other types of halo nuclei, we refer to
R. K. Gupta, S. Kumar, M. Balasubramaniam, G. M\"unzenberg, and W. Scheid, J. Phys. G \textbf{28}, 699 (2002);
W. Geithner \textit{et al.}, Phys. Rev. Lett. \textbf{101}, 252502 (2008); 
W. N\"ortersh\"auser \textit{et al.}, Phys. Rev. Lett. \textbf{102}, 062503 (2009). 

\bibitem{HF} S. S. Chandel and S. K. Dhiman, Phys. Rev. C \textbf{68}, 054320 (2003).

\bibitem{AS} M. Abramowitz and I. A. Stegun, {\it Handbook of Mathematical Functions} (Dover, New York, 1965).

\bibitem{GammaSource} A. Giulietti \textit{et al.}, Phys. Rev. Lett. \textbf{101}, 105002 (2008).

\bibitem{FLASH} W. Ackermann \textit{et al.}, Nat. Photon. \textbf{1}, 336 (2007); for current information, see http://flash.desy.de.

\bibitem{LCLS} L. Young \textit{et al.}, Nature \textbf{466}, 56 (2010); for current information, see http://lcls.slac.stanford.edu.

\bibitem{surface} A. Pukhov, Nature Phys. \textbf{2}, 439 (2006); 
B. Dromey \textit{et al.}, Nature Phys. \textbf{5}, 146 (2009)

\bibitem{LL} V. B. Berestetskii, E. M. Lifshitz, and L. P. Pitaevskii, {\it Relativistiv Quantum Theory} (Pergamon, New York, 1971). 

\bibitem{2s} See, e.g., M. S. Hussein, A. F. R. de Toledo Piza, O. K. Vorov, and A. K. Kerman, Phys. Rev. C \textbf{60}, 064615 (1999).

\bibitem{HHG} M. Klaiber, K. Z. Hatsagortsyan, C. M\"uller, and C.~H. Keitel, Opt. Lett. \textbf{33}, 411 (2008);
for an upcoming review, see M. C. Kohler, T. Pfeifer, K. Z. Hatsagortsyan, and C. H. Keitel, arXiv:1201.5094

\end{thebibliography}
\end{document}